\documentclass[preprint,superscriptaddress,pra]{revtex4}
\usepackage{amssymb,amsmath,graphicx,amscd,xcolor,dsfont}

\def\E{\mathbf{E}}

\begin{document}

\title{Subvacuum effects on light propagation}

\author{V. A. \surname{De Lorenci}}
\email{delorenci@unifei.edu.br}
\affiliation{Instituto de F\'{\i}sica e Qu\'{\i}mica,    Universidade Federal de Itajub\'a, \\
Itajub\'a, Minas Gerais 37500-903, Brazil}

\author{L. H. \surname{Ford}}
\email{ford@cosmos.phy.tufts.edu}
\affiliation{Institute of Cosmology, Department of Physics and Astronomy, \\
Tufts University, Medford, Massachusetts 02155, USA}

\begin{abstract}
Subvacuum effects arise in quantum field theory when a classically positive quantity, such as the local energy density, acquires a negative renormalized expectation value. 
Here we investigate the case of states of the quantized electromagnetic field with negative mean-squared electric field, and their effects on the propagation of light pulses in a 
nonlinear dielectric material with a nonzero third-order susceptibility. We identify two distinct signatures of the subvacuum effect in this situation. The first is an increase in the 
speed of the pulse, which is analogous to the superluminal light propagation in gravity which can arise from negative energy density. This increase in speed leads to a phase shift 
which might be large enough to observe. The second effect is a change in the frequency and power spectra of the pulse. We identify a specific measure of the modified spectra
 which can signal the presence of a negative mean squared electric field. These ideas are implemented in the particular example of a wave guide filled with a 
 nonlinear dielectric material. 
\end{abstract}

\maketitle

\section{Introduction}
A subvacuum effect in quantum field theory may be defined as a situation
where a classically positive quantity, such as the energy density or the squared electric field, acquires a negative expectation value when the formally divergent
part is subtracted. 
This can arise either in the presence of boundaries, such as the Casimir effect~\cite{BM69,SF02,SF05}, 
or in a nonclassical quantum state, such as a squeezed vacuum state~\cite{Caves} . 
In flat spacetime, the relevant operator is taken to be normal-ordered, so its vacuum expectation value vanishes,
and a locally negative expectation value represents a suppression of quantum fluctuations below the vacuum level, or a subvacuum effect. Negative
energy density and its possible effects in gravity have been extensively studied in recent years. This includes proving various quantum inequality
relations, which limit the magnitude and duration of subvacuum effects~\cite{ford78,ford91,fr95,fr97,flanagan97,FE98,pfenning01,fh05}.
One particularly striking effect of negative energy
density is its ability to increase the speed of light compared to its speed in the vacuum~\cite{PSW,Olum,VBL}. This can be viewed as a ``Shapiro time advance", 
and is the converse 
of the effect of positive energy, which produces the Shapiro time delay~\cite{Shapiro}. Of course, the gravitational effects of negative energy density are normally 
very small, so there is interest in finding analog effects in nongravitational systems which might be easier to observe in the laboratory. Some proposals
which have been made in the past include effects of vacuum fluctuation suppression on spin systems~\cite{fgo92} or on the decay rate of atoms in excited
states~\cite{FR11}. Another possibility involves the propagation of light in nonlinear materials, where electric field fluctuations could lead to fluctuations in the
speed of light~\cite{FDMS13,BDFS15,BDFR16}, and a negative mean squared electric field could increase the average speed of light in a material~\cite{BDF14}. 
Here we focus attention upon the latter effect
in a material with a nonzero third-order susceptibility, and will discuss two phenomenological signatures of the subvacuum effect. We consider a probe
wave packet propagating through a region with a negative mean-squared background electric field, and show that the subvacuum effect produces both a phase
advance of the packet, and particular features in its frequency and power spectra.

The wave equation of a probe field prepared in a single mode coherent state propagating in a nonlinear optical material is obtained in the next section, under certain conditions. 
It is shown that an applied background electric field prepared in a squeezed vacuum state couples to the nonlinearities of the medium and affects the motion of the probe field. 
Particularly, a phase shift occurs  whose magnitude may be large enough to be observable.  Section \ref{spectrum} deals with the influence of the quantum fluctuations 
of the background field on the spectrum of a propagating probe wave packet. Possible observable signatures related to subvacuum effects are presented in this section. 
In Sec.~\ref{guide} we explore these ideas in a model with a rectangular wave guide filled with a nonlinear dielectric. The results of the paper are summarized and discussed
 in Sec.~\ref{sec:final}.
Unless otherwise noted, we work in Lorentz-Heaviside units with $\hbar = c =1$, so $\epsilon_0 =1$.

\section{Phase shift}
\label{shift}
We start with the wave equation for an electric field in a nonlinear dielectric material~\cite{FDMS13},
\begin{equation}
\biggl(\nabla^2 - \frac{\partial^2}{\partial t^2}\biggr)E_i =  \frac{\partial^2}{\partial t^2}P_i\,,
\label{wave-eq}
\end{equation}
where we assume that $\nabla\cdot\E=0$, and $P_i$ represents the $i$-th component of the polarization vector, whose power series expansion in the electric field 
is given by (see, for example, Ref. \cite{boyd2008})
\begin{equation}
P_i = \chi_{ij}^{(1)} E_{j} + \chi_{ijk}^{(2)} E_{j}E_{k} 
+ \chi_{ijkl}^{(3)} E_{j}E_{k}E_{l} + \cdots  \,.
\label{polarization}
\end{equation}
Here summation on repeated indices is understood.
The coefficients $\chi_{ij}^{(1)}$ are the components of the linear susceptibility tensor, while  $\chi_{ijk}^{(2)}$ and $\chi_{ijkl}^{(3)}$ are the components of the 
second- and third-order nonlinear susceptibility tensors, respectively. We will be interested in centrosymmetric materials, for which the $\chi_{ijk}^{(2)}$ coefficients 
are identically zero. Additionally, if we specialize to the case where the electric field propagates in the $x$-direction and is linearly polarized in the $y$-direction, we obtain 
\begin{equation}
\biggl[\frac{\partial^2}{\partial x^2} - \epsilon\,\frac{\partial^2}{\partial t^2}\biggr]E =    \chi_{yyyy}^{(3)}\frac{\partial^2}{\partial t^2}E^3,
\label{wave-x}
\end{equation}
where we define $\epsilon =1+\chi^{(1)}_{yy}$. 
In order to keep the notation simpler, in what follows we will suppress the indices of the susceptibility coefficients. 

Suppose the electromagnetic field is prepared in a quantum state such that only two modes are excited, which will be called the probe (mode 1) and background (mode 2) fields. 
Let the probe field be a coherent state of amplitude $z$ described by the state vector $|z\rangle$, and the background field be a single-mode squeezed vacuum state defined by 
$|\zeta\rangle$. In general, $z$ and $\zeta$ are complex parameters, but we take them to be positive real numbers for simplicity, and set $\zeta = r$.
 Let us denote the state of the electromagnetic field as $|\psi\rangle = |z\rangle|\zeta\rangle$. Ignoring all other modes, the electric field operator can be expanded as 
$
\hat E = {\cal E}_1\hat a_1 + {\cal E}^*_1\hat a^\dagger_1+ {\cal E}_2\hat a_2  + {\cal E}^*_2\hat a^\dagger_2,
$ 
where ${\cal E}_i$ is the mode function for mode $i$,  and $\hat a_i$ and $\hat a_i^\dagger$ are the corresponding annihilation and creation operators, respectively. As $\langle\psi | \hat a_2 |\psi \rangle = \langle\zeta | \hat a_2 |\zeta \rangle = 0$,
the expectation value of the electric field operator when the system is prepared in the state $|\psi\rangle$ is  
\begin{equation}
{\cal E}_{c} = \langle\psi | \hat E|\psi \rangle  =    z \,({\cal E}_1  + {\cal E}^*_1 ) 
\end{equation}
where ${\cal E}_{c}$ can be viewed as the classical probe field. Similarly, the expectation value of normal-ordered $\hat E^3$ is given by
\begin{equation}
\langle \psi | :\hat E^3: |\psi \rangle  =  {{\cal E}_c}^3 + 3 {\cal E}_c \langle {E_q}^2 \rangle \,,
\end{equation}
where
\begin{equation}
\langle {E_q}^2 \rangle = \langle \zeta |:({\cal E}_2\hat a_2  + {\cal E}^*_2\hat a^\dagger_2 )^2:|\zeta\rangle = 
2\sinh r\left[|{\cal E}_2|^2 \sinh r - {\rm Re} \left({{\cal E}_2}^2 \right)\cosh r\right]
\end{equation}
is the mean value of the normal-ordered squared background electric field operator. The subvacuum effect arises when $\langle {E_q}^2 \rangle < 0$. Let the mode function for the background
field be a plane wave of the form ${\cal E}_2(x,t) = E_0\, {\rm e}^{i(k_b x -\Omega t)}$, so it is propagating in the $+x$-direction with angular frequency $\Omega$.
Then we have
\begin{equation}
\langle {E_q}^2 \rangle = 2 E_0^2 \, \sinh^2 r \{ 1 - \coth r \,\cos[2(k_b x - \Omega t)] \} \,.
\label{eq:Esq}
\end{equation}
Because $\coth r > 1$, we will have regions where $\langle {E_q}^2 \rangle < 0$ which travel at speed $\Omega/k_b$ in the $+x$-direction. The magnitude and duration of the 
$\langle {E_q}^2 \rangle < 0$ regions are constrained by quantum inequalities~\cite{fr97,FE98} of the form $\langle {E_q}^2 \rangle > -C/\tau^4 $, where $\tau$ is the temporal 
duration of the negative expectation value at one spatial point, and $C$ is a positive constant smaller than one. The essential physical content of these inequalities is that
there is an inverse relation between how negative  $\langle {E_q}^2 \rangle $ is, and how long the negative region can persist.
In the case of Eq.~(\ref{eq:Esq}), $\tau \alt 1/\Omega$
and states with $r \approx O(1)$ come closest to saturating the quantum inequality bounds~\cite{KF18}. A generalization of Eq.~(\ref{eq:Esq}) to a multimode
example is given in Sec.~\ref{guide}. In the Appendix, it is shown that this example satisfies a quantum inequality. 

The wave equation for the classical field ${\cal E}_c$ is obtained after using the above results in the quantum expectation value of Eq. (\ref{wave-x}), and becomes
\begin{equation}
\left[\frac{\partial^2}{\partial x^2} - \left(\epsilon +  3 \chi^{(3)}  \langle {E_q}^2\rangle\right) \frac{\partial^2}{\partial t^2}\right]{\cal E}_c 
= \chi^{(3)}\frac{\partial^2}{\partial t^2}{{\cal E}_c}^3 \,,
\label{probe-equation}
\end{equation}
where we have assumed that the classical probe field varies rapidly compared to the background field. Let us consider the case where the cubic term on the right hand
side of this equation may be neglected, so ${\cal E}_c$ approximately satisfies the linear equation 
\begin{equation}
\left[\frac{\partial^2}{\partial x^2} - \left(\epsilon +  3 \chi^{(3)}  \langle {E_q}^2\rangle\right) \frac{\partial^2}{\partial t^2}\right]{\cal E}_c  \approx 0\,.
\label{lin-equation}
\end{equation}
Then over a region small compared to the wavelength
of the background field, so $\langle {E_q}^2 \rangle $ is approximately constant, this equation describes waves with a phase velocity of $v_{\rm eff}$, where
\begin{equation}
v_{\rm eff} = \frac{1}{\sqrt{\epsilon +  3 \chi^{(3)}  \langle {E_q}^2\rangle}} \approx v_0\left(1 -  \frac{3\chi^{(3)}}{2\epsilon}  \langle {E_q}^2\rangle\right)\,.
\label{vpapprox}
\end{equation}
Here $v_0 = 1/\sqrt{\epsilon}$ is the phase velocity in the absence of the background field, and we have assumed  $\chi^{(3)}  |\langle {E_q}^2\rangle| \ll 1$. 
In regions where $\langle {E_q}^2 \rangle < 0$, we have that the speed of the probe field is increased, $v_{\rm eff} > v_0$, although $v_{\rm eff}$ is still less
than the speed of light in vacuum. This is the analog of the superluminal propagation of light in the presence of negative energy density in general relativity.

\begin{figure}
\includegraphics[scale=0.4]{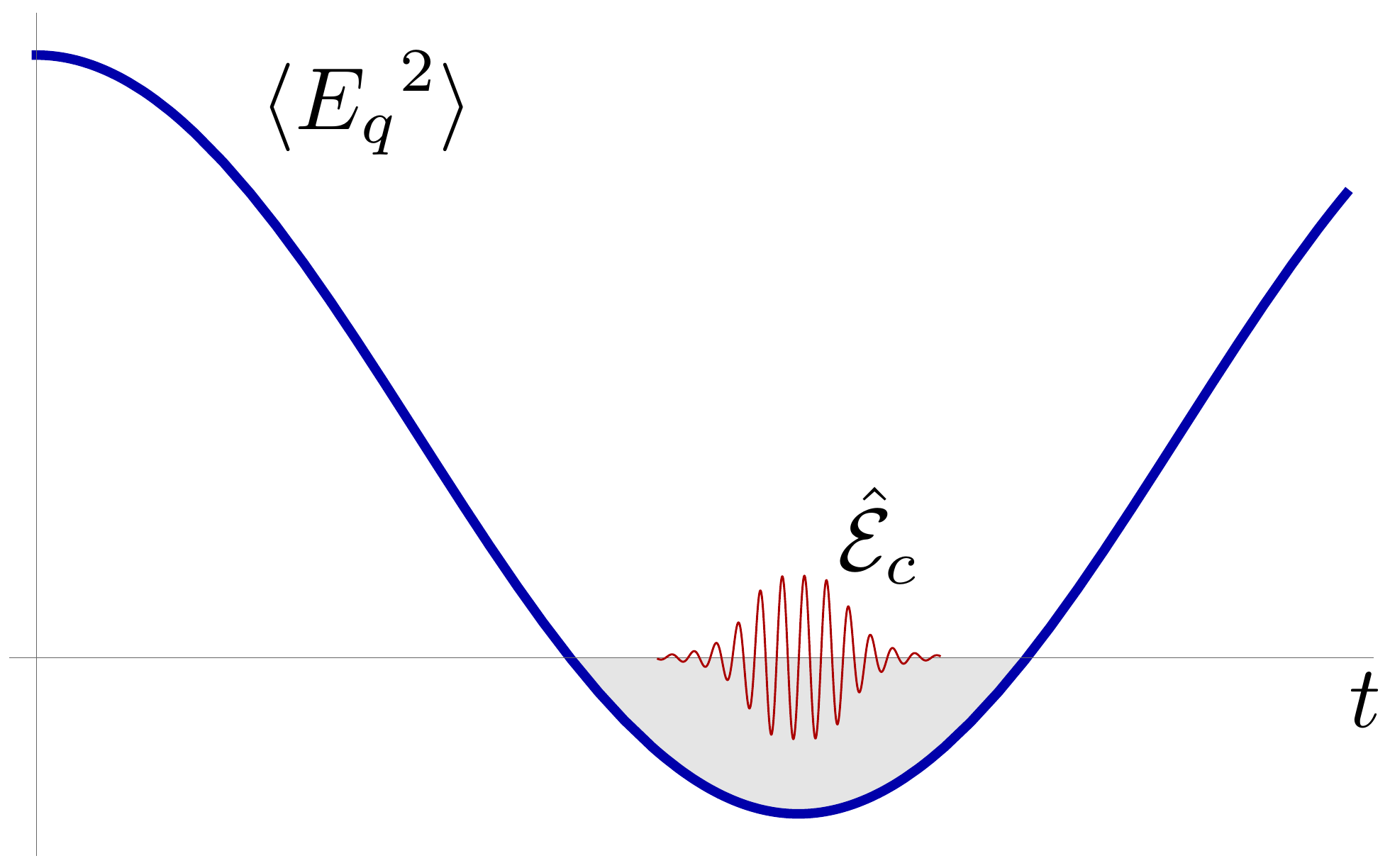}
\caption{Illustration of the probe wave packet in a region where $\langle {E_q}^2\rangle$ is negative.}
\label{fig0}
\end{figure}

Consider a wave packet solution of the linearized equation for ${\cal E}_c$ of the form
\begin{equation}
{\cal E}_c(x,t) = F(k x-\omega_0 t)\,  {\rm e}^{i(k x -\omega_0 t)}\,,
\end{equation}
where $F$ is an envelope function which varies more slowly than the exponential factor. In writing this form, we have assumed that dispersion can be ignored
over the bandwidth of the wave packet, so that both the phase and group velocities are approximately $ v_{\rm eff} =  \omega_0/k$. 
If $  \omega_0 \gg \Omega$,
we can select the envelope function so that the entire packet lies in a region where $\langle {E_q}^2 \rangle $ is both negative and approximately constant, as illustrated in Fig. \ref{fig0}. 
Further, let $\Omega/k_b \approx v_0$ so that this region moves at the same speed as does the  wave packet. This is possible if $\epsilon(\Omega)
\approx \epsilon(\omega_0)$. Note that this does not require that  $\epsilon(\omega)$ be constant over the entire interval from $\Omega$ to $\omega_0$.
If these conditions are satisfied, then the probe packet moves together with the region where $\langle {E_q}^2 \rangle < 0$. This will allow the phase shift effect of the background
 field on the wave packet to accumulate. 
 
 The effect with which we are dealing is quite different from the apparent superluminal propagation which can arise in a region of anomalous dispersion, as can occur
 in atomic vapors~\cite{Akulshin}. The latter effect occurs when the index of refraction is changing rapidly as a function of frequency and where the group and phase velocities can
 be very different from one another and from the signal velocity. This is also a frequency region where absorption is large, and pulse shapes can change rapidly. None of
 these features apply in the effect we are describing.

Note that as the wave packet enters a region where $\langle {E_q}^2 \rangle \not= 0$, the peak frequency $\omega_0$ is unchanged, but the wavenumber changes
from $k=k_0 = \omega_0/v_0$ to
\begin{equation}
k = \frac{\omega_0}{v_{\rm eff}} \approx   k_0 \left(1 +  \frac{3\chi^{(3)}}{2\epsilon}  \langle {E_q}^2\rangle\right)\,.
\end{equation}
 After a travel distance of $\Delta x =d$, this leads to a phase shift of the packet of
 \begin{equation}
\Delta \varphi = (k-k_0)\,d \approx 3 \frac{\chi^{(3)}}{2 \epsilon} \langle {E_q}^2\rangle \,k_0 \, d 
= 3\pi\frac{\chi^{(3)}}{\sqrt{\epsilon}} \langle {E_q}^2\rangle \left(\frac{d}{\lambda}\right)\,,
\end{equation}
where $\lambda = 2 \pi \sqrt{\epsilon}/ k_0$ is the wavelength of the probe field in the absence of the background field. To estimate the possible magnitude of this phase shift,
we may use
\begin{equation}
\Delta \varphi \approx \frac{1}{\sqrt{\epsilon}} \left(\frac{\chi^{(3)}}{3\times 10^{-19}{\rm m^2V^{-2} }}\right)   \left(\frac{\langle {E_q}^2\rangle}{1 (\mu{\rm m})^{-4} } \right)
 \left(\frac{d}{10{\rm m}}\right)\left(\frac{0.1\mu{\rm m}}{\lambda}\right) \,.
\label{dfi-estimates}
\end{equation}
The conversion to SI units is aided by noting that in our units $\epsilon_0  = 1$, which implies that $1V \approx 1.67\times 10^7 {\rm m}^{-1}$.

We can see from Eq.~(\ref{dfi-estimates}) that if a sufficiently large travel distance $d$ can be arranged, then a potentially observable phase shift could
result. Recall that this result was derived assuming that the nonlinear term in Eq.~(\ref{probe-equation}) can be neglected. This seems to require that
$ {{\cal E}_c}^2 \ll |\langle {E_q}^2 \rangle| $, and hence that $z \ll 1$. This in turn requires that the mean number of photons in the probe wave packet be
small compared to one. Nonetheless, it may be possible to build up an interference pattern with an extremely low count rate, but a long integration time.
Another possibility is that one might be able to take advantage of the dependence of the phase shift upon the background field even if the effect of the
nonlinear term is not negligible.

\section{Effects on the spectrum of the probe wave packet.}
\label{spectrum}
\subsection{Frequency spectrum of the probe field}

 Let 
 \begin{equation}
 f(x,t) = \frac{3\chi^{(3)}}{2\epsilon} \, \langle {E_q}^2\rangle(x,t) \,, 
 \label{eq:f}
 \end{equation}
 so that Eq.~(\ref{vpapprox}) becomes $v_{\rm eff} \approx v_0\, [1 - f(x,t)]$. We can write
 an approximate WKB solution of Eq.~(\ref{lin-equation}) as
 \begin{equation}
{\cal E}_c(x,t) = {\cal E}_0 {\rm e}^{ik_0 \left[x-v_0\left(1-f\right)t\right]} \approx {\cal E}_0 {\rm e}^{i\left(k_0 x-\omega_0t\right)} \left(1+ i \omega_0 f t \right)\,,
\label{wkb-expansion}
\end{equation}
where in the last step we assume $\left| f(x,t)\right| \omega_0 t \ll 1$. When the background field is described by a single plane wave mode state, as
in Eq.~(\ref{eq:Esq}), $f(x,t)$ has the form 
\begin{equation}
f(x,t) = \alpha +\beta\cos\left[2\left(k_b x-\Omega t\right)\right] \,,
\label{f}
\end{equation}
where $\alpha>0$ and $\beta$ are constants. In the case of a squeezed vacuum state, as in Eq.~(\ref{eq:Esq}), we have $|\beta| > \alpha$ and regions where 
$\langle {E_q}^2 \rangle < 0$. However,
the form of Eq.~(\ref{f}) could hold for a broader range of states, including more classical states where $\alpha> |\beta|$ and  $\langle {E_q}^2 \rangle > 0$ everywhere.
Here we show that there are features in the frequency spectrum of the probe wave packet which can distinguish these two cases.
 
\begin{figure}
\includegraphics[scale=0.4]{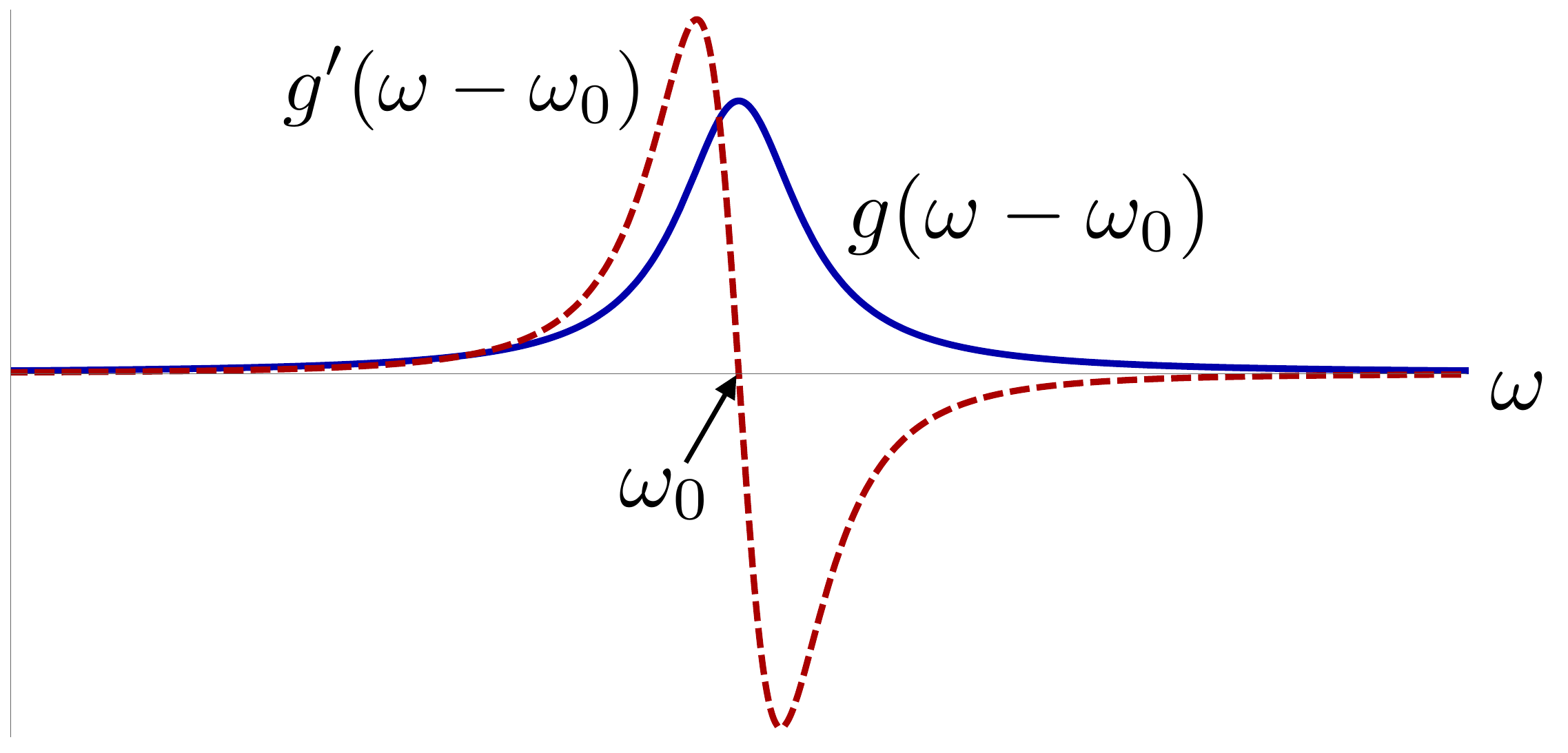}
\caption{The shapes of the function $g(\omega-\omega_0)$, described by a symmetric function centered at $\omega_0$, and its derivative, $g'(\omega-\omega_0)$,
are illustrated.}
\label{fig2}
\end{figure}

 The frequency spectrum can be defined by a Fourier transform of the probe electric field at a fixed spatial location of the form
 \begin{equation}
\hat{\cal E}_c(\omega) = \int_{-\infty}^{+\infty}{\rm e}^{i\omega t}{\cal E}_c(0,t) dt \,.
\label{fourier}
\end{equation}
Use Eqs.~(\ref{wkb-expansion}) and (\ref{f}) to find
\begin{eqnarray}
\hat{\cal E}_c(\omega) &=& 2\pi{\cal E}_0 \biggl\{ \delta(\omega-\omega_0) +\omega_0 \alpha\delta'(\omega-\omega_0)
\nonumber \\
&&+\frac{1}{2}\omega_0 \beta\bigl[\delta'(\omega-\omega_0+2\Omega) +\delta'(\omega-\omega_0-2\Omega) \bigr]\biggr\} \,,
\label{fourier-deltas}
\end{eqnarray}
where $\delta'(\omega) = d \delta(\omega)/d\omega$ is the derivative of a $\delta$-function and $\omega_0=k_0 v_0$. Equation~(\ref{fourier-deltas}) is the rather singular spectrum
associated with the monochromatic solution, Eq.~(\ref{wkb-expansion}). A more realistic solution would be a wave packet with a finite bandwidth. The spectrum
of such a solution can be obtained from Eq.~(\ref{fourier-deltas}) by replacing $\delta(\omega-\omega_0)$ by $g(\omega-\omega_0)$, a sharply peaked function
with finite width and unit area, such as a Lorentzian function. The expected functional form of $g$ and its first derivative are plotted in Fig.~\ref{fig2}.
In terms of the function $g$,   the frequency spectrum for a wave packet becomes 
\begin{eqnarray}
\hat{\cal E}_c(\omega) &=& 2\pi{\cal E}_0 \biggl\{ g(\omega-\omega_0) +\omega_0 \alpha g'(\omega-\omega_0)
\nonumber \\
&&+\frac{1}{2}\omega_0 \beta\bigl[g'(\omega-\omega_0+2\Omega) +g'(\omega-\omega_0-2\Omega) \bigr]\biggr\} \,.
\label{fourier-g}
\end{eqnarray}

\begin{figure}
\includegraphics[scale=0.33]{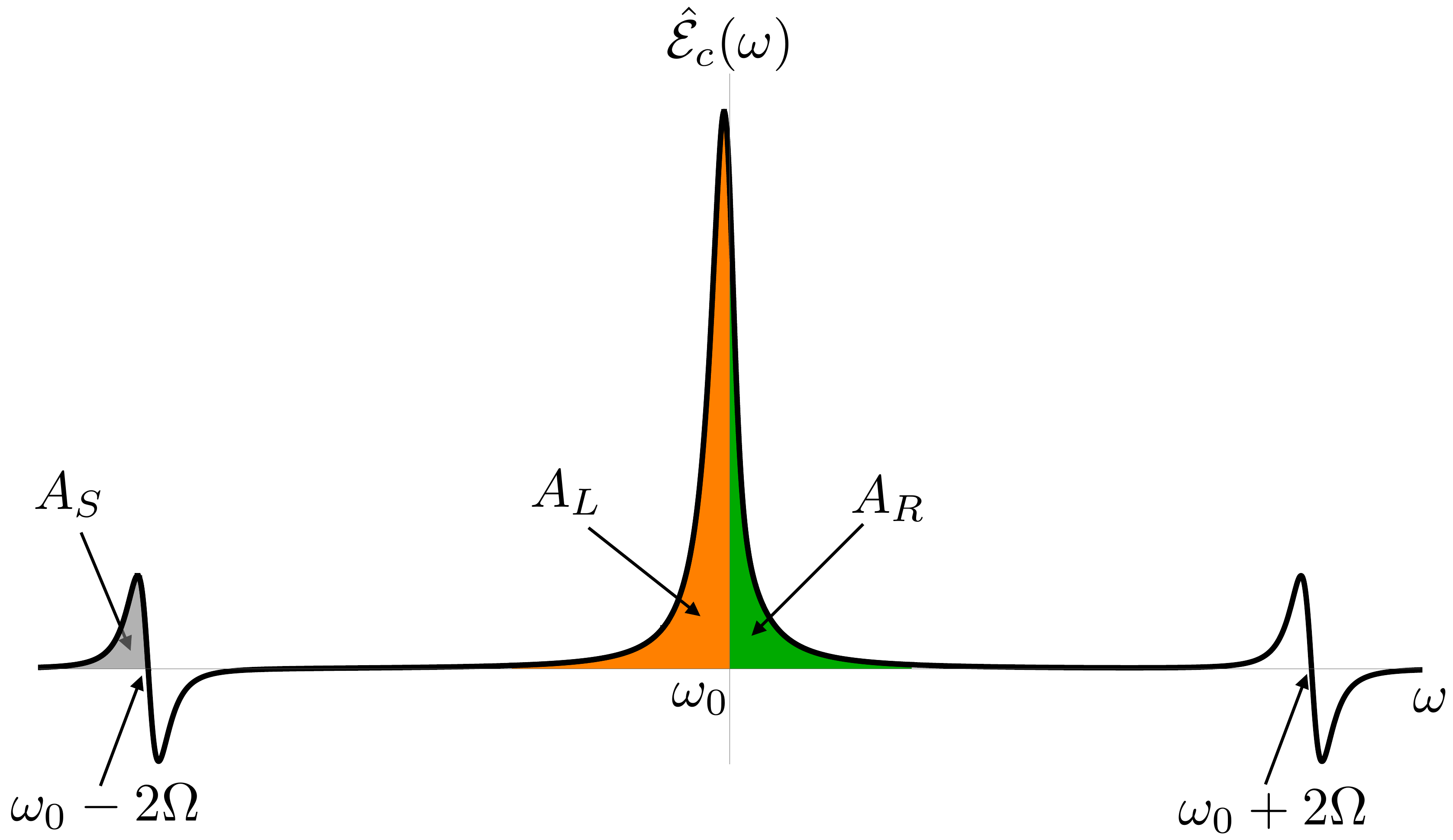}
\caption{The frequency spectrum of a probe wave packet is illustrated. The effect of the background field is to produce a shift of the central peak, and two
side bands at $\omega = \omega_0 \pm 2 \Omega$. The area of the central peak where $\omega < \omega_0$ is $A_L$ and that where $\omega > \omega_0$ 
is $A_R$. The area of one positive peak of a side band is $A_S$. }
\label{fig:freq-spectrum}
\end{figure}

Let us note some of the qualitative features of the spectrum in Eq.~(\ref{fourier-g}). When $\alpha=0$, there is a central peak at $\omega = \omega_0$. The term proportional
to $g(\omega-\omega_0)$ gives a symmetric contribution to this peak, but the term proportional to $\alpha$ produces a distortion which enhances the left
($\omega < \omega_0$) side and suppresses the right ($\omega > \omega_0$) side, if $\alpha > 0$. To leading order, this is a shift of the peak to the left. 
Let $A_L$ and $A_R$ be the areas of the left and right sides, respectively,
of $\omega = \omega_0$. The two terms in Eq.~(\ref{fourier-g}) proportional to $\beta$ produce side bands at $\omega = \omega_0 \pm 2 \Omega$. Each side band
consists of a positive and a negative peak. Let $A_S$ be the area of one positive side band peak. All of these features are illustrated in Fig.~\ref{fig:freq-spectrum}.
Note that $A_S$ may be written as
\begin{equation}
A_S = \left|\int_{-\infty}^{\omega_0 - 2 \Omega}  \hat{\cal E}_c(\omega) \, d\omega\right| =  \pi\, {\cal E}_0 \,\omega_0\, |\beta|\, g(0)\,
\label{eq:AS}
\end{equation}
if ${\cal E}_0  > 0$.
Similarly, the areas of the left and right sides of the central peak may be written as
\begin{equation}
A_L \approx \int_{\omega_0 - n\Delta\omega_p}^{\omega_0}  \hat{\cal E}_c(\omega) \, d\omega
\end{equation}
and
\begin{equation}
A_R \approx \int_{\omega_0}^{\omega_0 + n\Delta\omega_p}  \hat{\cal E}_c(\omega) \, d\omega\,,
\end{equation}
respectively. In the above integrals, $n$ is a number larger than one but $n\Delta\omega_p \ll 2\Omega$, where $\Delta\omega_p$ denotes the characteristic width 
of the function $g(\omega)$, i.e., the bandwidth of the probe field.  Both  $A_L$ and $A_R$ contain contributions from the symmetric  $g(\omega-\omega_0)$ term 
in Eq.~(\ref{fourier-g}). However,
if we take the difference, $A_L - A_R$, these contributions cancel and only the contribution from the term proportional to $\alpha$ remains, leading
to
\begin{equation}
A_L - A_R = 4 \pi {\cal E}_0\, \omega_0\, \alpha\, g(0)\,.
\label{eq:AL-AR}
\end{equation}
We may now use Eqs.~(\ref{eq:AS}) and (\ref{eq:AL-AR}) to write
 \begin{equation}
\frac{|\beta|}{\alpha} = \frac{4 A_S}{A_L - A_R} \,.
\label{eq:B/A} 
\end{equation}
Recall that when $|\beta| > \alpha$, a subvacuum effect is present in that  $\langle {E_q}^2 \rangle < 0$ somewhere, but when $|\beta| \leq \alpha$, the subvacuum effect 
is absent. Equation~(\ref{eq:B/A}) shows how detailed features in the frequency spectrum of the probe wave packet can distinguish between these two situations. 
Specifically, if $4 A_S > A_L - A_R $, then $\langle {E_q}^2 \rangle < 0$ somewhere.

If $g(\omega)$ has the form of a Lorentzian function, 
\begin{equation}
g(\omega) = \frac{\Delta\omega_p}{ \pi (\Delta\omega_p{}^2 + \omega^2)} \,,
\label{eq:Lorentzian}
\end{equation}
it follows that $A_L-A_R = 4{\cal E}_0\omega_0\alpha/\Delta\omega_p$.  As the central peak is described by $2\pi {\cal E}_0 g(\omega-\omega_0)$, 
its area $A_C = A_L + A_R$ can be approximated by $A_C \approx 2\pi {\cal E}_0$.  Hence, 
\begin{equation}
\frac{A_S}{A_C}=\frac{1}{2\pi}\left(\frac{\omega_0}{\Delta\omega_p}\right) |\beta|,
\label{assac}
\end{equation} 
and the asymmetry of the modified central peak can be described by means of
\begin{equation}
\frac{A_L-A_R}{A_C} = \frac{2}{\pi}\left(\frac{\omega_0}{\Delta\omega_p}\right)\alpha.
\label{almar}
\end{equation}
Here $\Delta\omega_p/\omega_0$ is the fractional line width of the probe field spectrum. 

In the previous section, we described the probe pulse as a localized wavepacket which tracks the region of $\langle {E_q}^2\rangle < 0$ in the background field, 
as illustrated in Fig. \ref{fig0}. Let $\tau_b = 2\pi/\Omega$ denote the period of the background field and  $\tau$ be the approximate temporal duration of 
the probe wavepacket, where $\tau < \tau_b$. This implies that the probe packet bandwidth, $\Delta\omega_p$, must to be larger than the background field angular frequency 
$\Omega$. Hence $(\Delta\omega_p/\omega_0) > (\Omega/\omega_0)$.
However, the present discussion of the effects of the background field on the probe field spectrum does not require such an assumption. Here we may assume that the probe field 
is approximately monochromatic, and nonzero in both $\langle {E_q}^2\rangle < 0$ and $\langle {E_q}^2\rangle > 0$ regions, which allows  $(\Delta \omega_p/\omega_0) \ll 1$.

\subsection{Power spectrum of the probe field}
\label{power}

In this subsection, we turn our attention to features of the probe pulse power spectrum, which is likely to be easier to measure than is $\hat{\cal E}_c(\omega)$,
the Fourier transform of the probe pulse electric field. 

The energy per unit area per unit time carried by the probe pulse in vacuum is given by the Poynting vector 
$\vec{S} =  \vec{\cal E}_c \times \vec B_c$.  It follows that $|\vec{S}| = |\vec{\cal E}_c(x,t)|^2= |{\cal E}_c(x,t)|^2$. 
Setting $x=0$, the total energy per unit area in the probe pulse can be written as
\begin{equation}
u = \int_{-\infty}^{\infty} |{\cal E}_c(0,t)|^2 dt = \frac{1}{2\pi}\int_{-\infty}^{\infty} |\hat{\cal E}_c(\omega)|^2 d\omega = \int_{-\infty}^{\infty} P(\omega) d\omega,
\label{uT}
\end{equation}
where Parseval's theorem was used in the second equality. In the above result we defined the power spectrum of the probe field as $P(\omega)=(1/2\pi)|\hat{\cal E}_c(\omega)|^2$. 
Using Eq.~(\ref{fourier-g}), we find
\begin{eqnarray}
P(\omega) &=&  2\pi{\cal E}_0^2 \biggl\{ g(\omega-\omega_0)^2 +2\omega_0\alpha g(\omega-\omega_0)g'(\omega-\omega_0)+\omega_0^2\alpha^2 g'(\omega-\omega_0)^2
\nonumber \\
&&+\frac{1}{4}\omega_0^2 \beta^2\bigl[g'(\omega-\omega_0+2\Omega)^2 +g'(\omega-\omega_0-2\Omega)^2 \bigr]\biggr\} \,,
\label{P(w)}
\end{eqnarray} 
where we assume that $2 \Omega \gg \Delta \omega_p$, and thus cross terms involving $g'(\omega-\omega_0\pm 2\Omega)$ can be neglected. 
Recall that $\Delta \omega_p$ is the approximate width of the central peak in Fig.~\ref{fig:freq-spectrum}, while $2\Omega$ is the separation of the sidebands from the central peak.
The energy per unit area contained in the frequency interval $\omega_1 \le \omega \le \omega_2$ becomes $\int_{\omega_1}^{\omega_2} P(\omega) d\omega$. 

\begin{figure}
\includegraphics[scale=0.3]{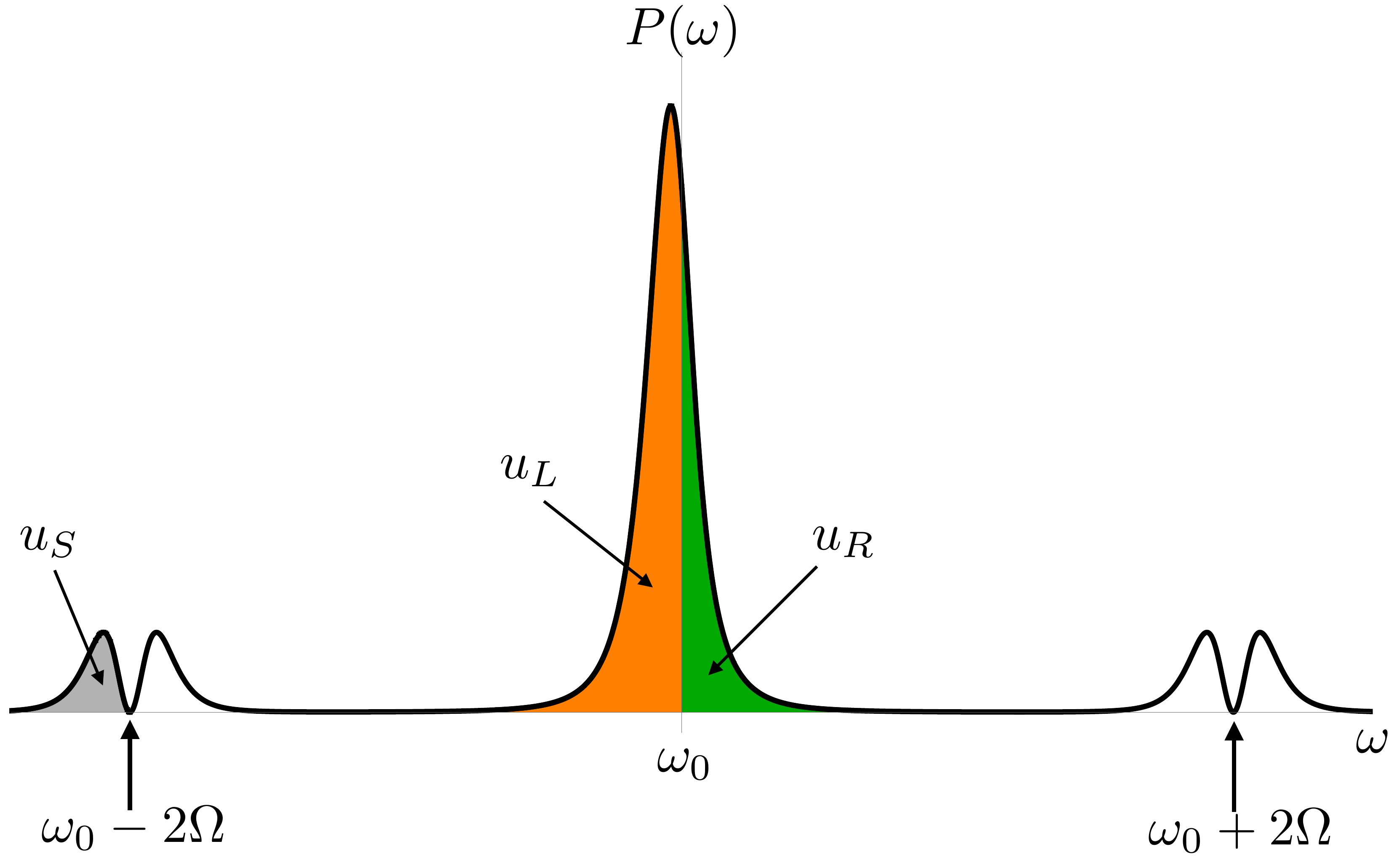}
\caption{The power spectrum of the probe wave packet is here illustrated. The effect of the background field is to produce a shift of the central peak, and two double peaked
side bands at $\omega = \omega_0 \pm 2 \Omega$. The area of the central peak where $\omega < \omega_0$ is $u_L$ and that where $\omega > \omega_0$ 
is $u_R$. The area of one positive peak of a side band is $u_S$.}
\label{fig4}
\end{figure}

The power spectrum, $P(\omega)$, has several features which are similar to those of the frequency  spectrum, which was illustrated in Fig. \ref{fig:freq-spectrum}. 
Both spectra have the central peak displaced to the left of $\omega_0$, and  sidebands at $\omega = \omega_0 \pm 2 \Omega$. The power spectrum is
illustrated in Fig.~\ref{fig4} for the case that $g(\omega)$ is the Lorentzian function given in Eq.~(\ref{eq:Lorentzian}).

The energy corresponding to the left and right sides of the central peak are given respectively by 
\begin{eqnarray}
u_L = \int_{\omega_0-n\Delta\omega_p}^{\omega_0} P(\omega) d\omega = 
\frac{{\cal E}_0^2}{2\Delta\omega_p}\left[1+\frac{4\omega_0\alpha}{\pi\Delta\omega_p}+\frac{\omega_0^2\alpha^2}{2\Delta\omega_p{}^2}\right],
\label{uL} 
\\
u_R = \int_{\omega_0}^{\omega_0+n\Delta\omega_p} P(\omega) d\omega = 
\frac{{\cal E}_0^2}{2\Delta\omega_p}\left[1-\frac{4\omega_0\alpha}{\pi\Delta\omega_p}+\frac{\omega_0^2\alpha^2}{2\Delta\omega_p{}^2}\right].
\label{uR}
\end{eqnarray}
Here we have used Eqs.~(\ref{eq:Lorentzian}) and (\ref{P(w)}).
The asymmetry in the power spectrum about the central peak leads to the following fractional asymmetry in the energy distribution associated with the probe pulse, 
\begin{equation}
\frac{u_L-u_R}{u_C} = \frac{4}{\pi}\left(\frac{\omega_0}{\Delta\omega_p}\right)\alpha
\label{uR2}
\end{equation}
where  $u_C \doteq {\cal E}_0^2/\Delta\omega_p$ is the energy in the central peak in the limit that $\alpha =0$. This expression is the analog of Eq.~(\ref{almar}).

Finally, the area of a sideband depicted in Fig. \ref{fig4}. The energy carried by this part of the probe pulse is 
\begin{equation}
u_S= \int_{-\infty}^{\omega_0-2\Omega} P(\omega) d\omega,
\label{uS}
\end{equation} 
which leads to the fractional energy in the side band,
\begin{equation}
\frac{u_S}{u_C} = \frac{1}{16}\left(\frac{\omega_0}{\Delta\omega_p}\right)^2\beta^2.
\label{uSfrac}
\end{equation}
In contrast to its analog for the frequency spectrum, Eq.~(\ref{eq:B/A}), the above expression is quadratic both in ${\omega_0}/{\Delta\omega_p}$ and in $\beta$. Thus,
if $(\omega_0/\Delta\omega_p) |\beta| \ll 1$, measurement of $u_S/u_0$ could be difficult. In this case, the frequency spectrum might become a better probe of
the presence of a subvacuum effect.

\section{Mode functions in a rectangular wave guide}
\label{guide}

Let us now consider the specific case where  an  electromagnetic wave  propagates in a rectangular wave guide whose cross section has dimensions 
$a$ and $b$, respectively. This example should provide a reasonable estimate of the magnitude of the effects expected in an optical fiber, but is
easier to compute in detail.
Suppose the boundaries are perfectly conducting and the inner region is filled with a  dielectric 
material whose linear contribution to the electric permittivity is given by $\epsilon$. 
Solutions for the transverse electric (TE)  modes propagating along the length of the wave guide can be expressed as~\cite{Zangwill}
\begin{eqnarray}
B_z &=& B_0\, \cos\left(\frac{\pi m x}{a}\right)\, \cos\left(\frac{\pi n y}{b}\right)\, {\rm e}^{i\phi} \, , \nonumber \\
B_x &=& \frac{i k}{\gamma^2} \; \frac{\partial B_z}{\partial x} \, , \qquad B_y = -\frac{i k}{\gamma^2} \; \frac{\partial B_z}{\partial y} \, , \nonumber \\
E_x &=& \frac{\omega}{k} \, B_y\,, \qquad E_y = - \frac{\omega}{k} \, B_x\,, \qquad E_z = 0 \,.
\end{eqnarray}
In a change from the notation of previous sections, the wave is now propagating in the $+z$-direction.
Here  $\phi = k z - \omega t$,   $k$ is the wavenumber, and $m$ and $n$ are positive integers, 
\begin{equation}
\gamma = \sqrt{\left(\frac{\pi m}{a}\right)^2 + \left(\frac{\pi n}{b}\right)^2 }\,,
\label{eq:gamma}
\end{equation}
and 
\begin{equation}
\omega = \frac{1}{\sqrt{\epsilon}} \, \sqrt{\gamma^2 + k^2} 
\label{eq:omega}
\end{equation}
is the angular frequency of the mode.

These modes are normalized so that 
\begin{equation}
\frac{1}{2} \int (\epsilon |{\bf E}|^2 + |{\bf B}|^2) d^3x = \frac{1}{2} \omega 
\end{equation}
is the zero point energy of a single mode. This determines the constant $B_0$ to be
\begin{equation}
B_0 = \gamma\, \sqrt{\frac{2}{a\, b\, L\, \epsilon\, \omega}}\,,
\end{equation}
where we impose periodic boundary conditions of length $L$ in the $z$-direction. 

Now we generalize the treatment in Sec.~\ref{shift}, and assume that the quantized electric field is prepared in a multimode squeezed vacuum state. We assume that
the excited modes are associated with specific values of $m$ and of $n$, but a finite bandwidth of the wavenumber $k$. The mean-squared electric field is now given by
\begin{eqnarray}
\langle {E_q}^2\rangle &=& \frac{4}{a\, b\, L\, \gamma^2 }\, \left[ \left(\frac{\pi m}{a}\right)^2\,  \sin^2\left(\frac{\pi m x}{a}\right)\, \cos^2\left(\frac{\pi n y}{b}\right) + 
 \left(\frac{\pi n}{b}\right)^2\, \cos^2\left(\frac{\pi m x}{a}\right)\, \sin^2\left(\frac{\pi n y}{b}\right)\, \right]  \nonumber \\
&\times&  \sum_k \frac{\omega}{\epsilon}  \sinh r_k \left[\sinh r_k - \cosh r_k \cos (kz-\omega t)\right],
\label{eq2int0}
\end{eqnarray}
where $r_k$ denotes the squeeze parameter corresponding to mode $k$.  In the Appendix, we show that Eq.~(\ref{eq2int0}) satisfies a quantum inequality bound. 
Next we average $\langle {E_q}^2\rangle$ over the cross section of the wave guide, and
let $\sum_k \rightarrow (L/2\pi)\int dk$, which leads to
\begin{equation}
\langle {E_q}^2\rangle = \frac{1}{2\pi\, a\,b} \, \int dk\, \frac{\omega}{\epsilon}\,  \sinh r_k \left[\sinh r_k - \cosh r_k \cos (kz-\omega t)\right].
\label{eq2int}
\end{equation}
As before, this quantity will be negative when $\cos (kz-\omega t) \approx 1$. We assume that  the squeeze parameter $r_k$ is different from 
zero only within a finite bandwidth $\Delta k$ about $k=  \sqrt{\epsilon}\, \Omega$. Here the excited modes of the background field are in a finite  angular frequency band
peaked at $\omega = \Omega$. This leads to the value of $\langle {E_q}^2\rangle$ near its minimum of
\begin{equation}
\langle {E_q}^2\rangle_{\bf min} \approx -\frac{\Omega\, \Delta k }{2 \pi \,a\,b\, \epsilon} \; \sinh r_k \left(\cosh r_k - \sinh r_k \right) \,.
\label{eq2dk}
\end{equation}

Let us now examine the effects of the $\langle {E_q}^2\rangle <0$ regime on the probe pulse spectrum. The wave equation is the same as Eq.~(\ref{lin-equation}),
but exchanging the variable $x$ by $z$. The WKB solution for ${\cal E}_c(z,t)$ is now given by Eq. (\ref{wkb-expansion}) with  
\begin{equation}
f(z,t) = - G\, \sinh r_k \left(\cosh r_k - \sinh r_k \right)\,,
\label{fz}
\end{equation}
where
\begin{equation}
G = \frac{3\chi^{(3)}\Omega \,\Delta k }{4\pi \,a\,b\,\epsilon^2} \,.
\label{eq:G}
\end{equation}
From Eq.~(\ref{fz}), we identify the  coefficients appearing in Eq.~(\ref{f}) to be $\alpha = G \sinh^2 r_k $ and  $|\beta| =  G \sinh r_k\cosh r_k$.
Note that both $\alpha$ and $|\beta|$ grow exponentially as $r_k$ increases, but their difference remains finite
\begin{equation}
|\beta| - \alpha = G \sinh r_k \left(\cosh r_k - \sinh r_k \right)=   G (1-{\rm e}^{-2r_k})/2 \approx \frac{1}{2} G \, \qquad r_k\gtrsim 1 \,.
\label{eq:beta-alpha}
\end{equation}
It is this difference which determines the magnitude of the subvacuum effect, which is our principal interest. Thus, we consider the case
where $r_k$ is of order one in the region where it is nonzero. Then, in order of magnitude, we have
\begin{equation}
|\beta| \approx  \alpha \approx |\beta|-\alpha \approx G\,.
\label{eq:beta-app}
\end{equation}

The quantitiy  $(\omega_0/\Delta \omega_p) \, \alpha  \approx (\omega_0/\Delta \omega_p) \, |\beta|$, which appears in 
Eqs.~(\ref{assac}), (\ref{almar}), (\ref{uR2}), and (\ref{uSfrac}) may be estimated from Eqs.~(\ref{eq:G}) and (\ref{eq:beta-app}) as 
\begin{equation}
 \left(\frac{\omega_0}{\Delta \omega_p}\right)\, |\beta|  \approx 
 \frac{1.0\times10^{-8}}{\epsilon^3} \left(\frac{\chi^{(3)}}{3\times10^{-19}{\rm m^{2}V^{-2}}}\right) \left(\frac{\Delta k}{\Omega}\right) \left(\frac{\omega_0}{\Delta \omega_p}\right)\
 \left(\frac{1\mu{\rm m}}{\lambda_b}\right)^2 \frac{(1\mu{\rm m})^2}{a\,b},
\label{beta-est}
\end{equation}
Recall that fractional asymmetry of both the frequency and power spectra, as well as the fractional area of the sidebands of the frequency spectra are all of the order
of the quantity in Eq.~(\ref{beta-est}), while the  fractional area of the sidebands of the power spectrum is proportional to its square.
Note that this quantity is proportional to the fractional bandwidth of the background field modes, ${\Delta k}/{\Omega}$, which need not be especially small, and
inversely proportional to the fractional bandwidth of the probe field wavepacket,  ${\Delta\omega_p}/{\omega_0}$,  which can be very small. 
Hence, if we were to set  $\omega_0/\Delta\omega_p \approx {\cal O}(10^{8})$, which is far from the narrowest possible line, we could have  fractional results 
described by Eqs. (\ref{assac}) and (\ref{almar}) of order 1. However, the magnitude of $\omega_0/\Delta\omega_p$ is limited by the approximation used in the WKB solution given by 
Eq.~(\ref{wkb-expansion}), that requires $\omega_0 f t < 1$, and so $t < 1/(\omega_0f) \approx 1/(\omega_0\alpha) \approx 1/(\omega_0 |\beta|)$. Thus, the Taylor-expanded  solution 
for ${\cal E}_c(x,t)$ is restricted to times obeying this condition, which puts a lower band on the bandwidth $\Delta\omega_p$ of $\Delta\omega_p \gtrsim 1/t > \omega_0\alpha$  
(or $1/t > \omega_0 |\beta|$). Hence, we need both $(\omega_0/\Delta\omega_p)\alpha$ and $(\omega_0/\Delta\omega_p) |\beta|$ to be  smaller than unity. As a consequence,
the various features of the frequency and power spectra, $A_S/A_C$,  $(A_L-A_R)/A_C$,   $(u_L-u_R)/u_C$,  and $u_S/u_C$, given in Eqs.~(\ref{assac}), (\ref{almar}),
(\ref{uR2}), and  (\ref{uSfrac}), respectively, all have to be small compared to one. However, if the spectra can be measured to sufficient accuracy, it may be possible
to confirm the existence of the subvacuum effect.

\section{Summary and Discussion}
\label{sec:final}

In this paper, we have explored some consequences of a negative mean-squared electric field as an example of a subvacuum effect, one where quantum fluctuations
are suppressed below the vacuum level. We consider the quantized electric field in a squeezed vacuum state, where $\langle {E_q}^2 \rangle < 0$ in some regions.
This forms a background field which can increase the speed of propagation of a probe pulse in a nonlinear material with nonzero third-order susceptibility. This is an analog
of the effect of negative energy density in general relativity, which can increase the effective speed of light. 

Our primary concern is a search for systems where the increased light speed, or related effects, in a nonlinear dielectric might be observable. The fractional increase in
light speed is typically both very small, perhaps of the order of $10^{-9}$, and transient.  However, in Sec.~\ref{shift}, we discussed the possibility that the probe
pulse wavepacket can travel with the region of negative mean-squared electric field, and hence the phase shift from the speed increase might accumulate to a measurable
magnitude. In Sec.~\ref{spectrum}, we discussed the effects of a region where $\langle {E_q}^2 \rangle < 0$ on both the frequency spectrum and power spectrum
of the probe pulse. We showed that the details of these spectra can carry information about whether the pulse has travelled through a region where 
$\langle {E_q}^2 \rangle < 0$. These ideas were developed further in Sec.~\ref{guide} in the context of pulses in a rectangular wave guide. This example allowed us to give
some estimates of the magnitudes of the spectra features produced by a negative mean-squared electric field, which indicate that they might be observable.

\begin{acknowledgments}
We would like to thank Peter Love for comments on the manuscript.
This work was supported in part by the Brazilian agency {\em Conselho Nacional de Desenvolvimento Cient\'{\i}fico e Tecnol\'ogico}  (CNPq, Grant 302248/2015-3) and 
by the National Science Foundation (Grant PHY-1607118).
\end{acknowledgments}

\appendix
\section{} 
\label{appendix}

In this appendix, we show that $\langle {E_q}^2 \rangle$ in a rectangular wave guide satisfies the quantum inequality constraint derived in Refs.~\cite{fr97,FE98}. 
This constraint is a lower bound of the form
\begin{equation}
\langle {E_q}^2 \rangle \geq -\frac{C}{\tau^4} 
\label{eq:QI}
\end{equation}
where $0 < C \alt 1$ and $\tau$ is the temporal duration of the negative mean squared  electric field at a given point in space. The inequalities derived in 
Refs.~\cite{fr97,FE98} generally require $\langle {E_q}^2 \rangle$ to be averaged in time with a sampling function of characteristic width $\tau$. However, here 
we have a finite bandwidth of excited modes and can
show that the results found in Sec.~\ref{guide} satisfy Eq.~(\ref{eq:QI}) without the need for time averaging. 

We begin with Eq.~(\ref{eq2int0}), and note that $\langle {E_q}^2 \rangle$ has its minimum value as a function of time when $\cos (kz-\omega t) = 1$ and that 
the magnitudes of sine and cosine functions are bounded above by unity, so
\begin{equation}
\langle {E_q}^2 \rangle \geq  -\frac{1}{\pi\, a\,b} \, \int dk \; \frac{\omega}{\epsilon}  \,.
\label{eq:QI2}
\end{equation}
Here we have used Eq.~(\ref{eq:gamma}) and the fact [See Eq.~(\ref{eq:beta-alpha}).]  that $ \sinh r_k \left(\sinh r_k - \cosh r_k \right) \geq - \frac{1}{2}$.
This bound is similar to the estimates given in Sec.~\ref{guide} , although here we do not assume a spatial average over the wave guide cross section.
As before, we assume a bandwidth of excited modes in a wave number range $\Delta k$ peaked near $k = \bar{k}$, and hence in angular frequency near
$\omega = \Omega = \sqrt{(\gamma^2 + \bar{k}^2)/\bar{\epsilon})}$, where $\bar{\epsilon} = \epsilon(\Omega)$.  
 Because $\Delta k < \bar{k} < \Omega \sqrt{\bar{\epsilon}}$, we may write
\begin{equation}
\langle {E_q}^2 \rangle \geq  -\frac{\Omega^2}{\pi\, a\,b\, \sqrt{\bar{\epsilon}}} \,.
\label{eq:QI3}
\end{equation}

We have $\gamma  \geq \pi m/a$ and  $\gamma  \geq \pi n/b$. Because both $m$ and $n$ are positive integers, we
have
 \begin{equation}
\frac{\Omega^2}{\pi\, a\,b\, \sqrt{\bar{\epsilon}}} \leq \frac{ \gamma^2\,  \Omega^2}{\pi^3\, m\,n\, \sqrt{\bar{\epsilon}}} \leq 
\frac{ \gamma ^2\,  \Omega^2}{\pi^3 \, \sqrt{\bar{\epsilon}}} <  \frac{ \sqrt{\bar{\epsilon}}\, \Omega^4}{\pi^3}\,.
\end{equation}
Finally, because $\bar{\epsilon}$ is of order one, and $\tau$ is of order $1/\Omega$, we obtain the quantum inequality bound, Eq.~(\ref{eq:QI}).

\end{document}